\documentclass[pdftex]{article}

\usepackage[nonatbib,final]{neurips_2022}

\usepackage[utf8]{inputenc} 
\usepackage[T1]{fontenc}    
\usepackage{url}            
\usepackage{booktabs}       
\usepackage{amsfonts}       
\usepackage{nicefrac}       
\usepackage{microtype}      
\usepackage{xcolor}         
\usepackage{wrapfig}
\usepackage{tikz}
\usepackage{tikz-feynhand}
\usepackage[subrefformat=parens]{subcaption}
\usepackage{booktabs}
\usepackage{multirow}
\usepackage{pdfpages}
\usepackage{hyperref}       

\title{Decay-aware neural network for event classification in collider physics}

\author{%
  Tomoe Kishimoto\\
  Computing Research Center, High Energy Accelerator Research Organization\\
  1-1 Oho, Tsukuba, Ibaraki, Japan \\
  International Center for Elementary Particle Physics, The University of Tokyo \\
  Institute for AI and Beyond, The University of Tokyo \\
   7-3-1 Hongo, Bunkyo-ku, Tokyo, Japan \\ 
  \texttt{tomoe.kishimoto@kek.jp} \\
   \And
   Masahiro Morinaga \\
   International Center for Elementary Particle Physics, The University of Tokyo \\
     Institute for AI and Beyond, The University of Tokyo \\
   7-3-1 Hongo, Bunkyo-ku, Tokyo, Japan \\ 
   \texttt{morinaga@icepp.s.u-tokyo.ac.jp} \\
      \And
   Masahiko Saito \\
   International Center for Elementary Particle Physics, The University of Tokyo \\
     Institute for AI and Beyond, The University of Tokyo \\
   7-3-1 Hongo, Bunkyo-ku, Tokyo, Japan \\ 
   \texttt{saito@icepp.s.u-tokyo.ac.jp} \\
         \And
   Junichi Tanaka \\
   International Center for Elementary Particle Physics, The University of Tokyo \\
     Institute for AI and Beyond, The University of Tokyo \\
   7-3-1 Hongo, Bunkyo-ku, Tokyo, Japan \\ 
   \texttt{jtanaka@icepp.s.u-tokyo.ac.jp} \\
}

\begin{document}

\maketitle

\begin{abstract}
The goal of event classification in collider physics is to distinguish signal events of interest from background events to the extent possible to search for new phenomena in nature. We propose a decay-aware neural network based on a multi-task learning technique to effectively address this event classification. The proposed model is designed to learn the domain knowledge of particle decays as an auxiliary task, which is a novel approach to improving learning efficiency in the event classification. Our experiments using simulation data confirmed that an inductive bias was successfully introduced by adding the auxiliary task, and significant improvements in the event classification were achieved compared with boosted decision tree and simple multi-layer perceptron models.
\end{abstract}

\section{Introduction}
In collider physics experiments, a large number of events\footnote{The term ``event'' corresponds to ``image'' in image classification.} are produced from particle collisions using high-energy accelerators, such as the Large Hadron Collider~\cite{Evans_2008}. For data analysis, event classification becomes crucial as it enables an effective separation of interesting signal events from background events. Machine learning, such as boosted decision trees (BDT), has a long history in collider physics~\cite{Albertsson_2018, Radovic2018MachineLA}; deep learning (DL) is also widely used to enhance the performance of event classification. DL can provide significant discrimination power by utilizing its huge parameter space; however, a large amount of data is  required to maximize its performance. A multi-task learning technique is a feasible approach to improving learning efficiency by introducing an inductive bias. Multiple related tasks are trained simultaneously while using a shared representation in multi-task learning, which may result in better performance than training the tasks individually. 

A decay-aware neural network based on the multi-task learning technique is proposed in this paper to efficiently address the event classification problem. In the proposed model, the main event classification task and an auxiliary task are trained simultaneously, where the domain knowledge of particle decays, as shown in Figure~\ref{fig:feyn}, is learned as the auxiliary task. In collider physics experiments, heavy particles of interest, such as Higgs boson~($h$), immediately decay to other particles and they are not observed by the detectors. Therefore, the auxiliary task based on particle decay knowledge is expected to provide a good inductive bias because the features of these heavy particles can be extracted efficiently from the observed particles. 

In this study, we perform experiments using simulation data to evaluate the decay-aware neural network. Furthermore, we also present the details of datasets and the proposed model structure. The remainder of this paper is organized as follows. Section~2 describes the related works, including our novelty. Section~3 summarizes the datasets used in this study. Section~4 provides details of the proposed model. Section~5 presents the experimental results. Finally, Section~6 concludes the paper.

\section{Related work}
A previous study reported that a multi-layer perceptron (MLP) model outperformed traditional BDT by identifying powerful features for the event classification~\cite{Baldi:2014kfa}. Graph neural network (GNN)~\cite{https://doi.org/10.48550/arxiv.1806.01261} architecture has also been reported to enhance the event classification performance~\cite{Abdughani_2019,Ren_2020}. Although we use this GNN architecture in our proposed model to extract graph-level and edge-level features, in contrast to these previous studies, our study has the following contributions:

\begin{itemize}
\setlength{\leftskip}{-0.5cm}
\item The multi-task learning is introduced to improve event classification performance. The rich structure of GNN enables solving multiple tasks simultaneously. The event classification and auxiliary tasks are implemented as the graph-level and edge-level classifications, respectively. 
\item Domain knowledge about particle decay is learned in the auxiliary task. The labels of auxiliary tasks are prepared by counting the decay points between the observed objects.
\end{itemize}

\section{Datasets}
\begin{figure}[htbp]
\begin{minipage}{0.33\hsize}
\centering
\begin{tikzpicture}[scale=0.85]
\begin{feynhand}

\vertex [particle] (i1) at (-2.5,1.25) {};
\vertex [particle] (i2) at (-2.5,-1.25) {};

\vertex [particle] (f1) at (2.5,1.25) {$\ell^{+}$};
\vertex [particle] (f2) at (2.5,0.75) {$\nu$};
\vertex [particle] (f3) at (2.5,0.25) {$b_{1}$};
\vertex [particle] (f4) at (2.5,-0.25) {$\bar{b}_{2}$};
\vertex [particle] (f5) at (2.5,-0.75) {$q_{1}$};
\vertex [particle] (f6) at (2.5,-1.25) {$\bar{q}_{2}$};

\vertex (v1) at (-1.5,0);
\vertex (v2) at (-0.5,0);
\vertex (v3) at (0.5,0.5);
\vertex (v4) at (1.5,1);
\vertex (v5) at (1.5,0.);
\vertex (v6) at (0.5,-0.5);

\propag [fer] (i1) to (v1);
\propag [antfer] (i2) to (v1);
\propag [sca] (v1) to [edge label=$H^{0}$] (v2);
\propag [sca] (v2) to [] (v3);
\propag [boson] (v3) to [edge label=$W^{+}$] (v4);
\propag [sca] (v3) to [edge label=$h$] (v5);
\propag [antfer] (v4) to (f1);
\propag [fer] (v4) to (f2);
\propag [fer] (v5) to (f3);
\propag [antfer] (v5) to (f4);
\propag [boson] (v2) to [edge label=$W^{-}$] (v6);
\propag [fer] (v6) to (f5);
\propag [antfer] (v6) to (f6);

\node at (-0.2,0.5) {$H^{+}$};
\node at (-0.2,-1.3) [font=\fontsize{0.9em}{0.9em}\selectfont] {(a) 2HDM signal process};

\end{feynhand}
\end{tikzpicture}

\end{minipage}
\begin{minipage}{0.33\hsize}
\centering
\begin{tikzpicture}[scale=0.85]
\begin{feynhand}

\vertex [particle] (i1) at (-2.5,1.25) {};
\vertex [particle] (i2) at (-2.5,-1.25) {};

\vertex [particle] (f1) at (2.5,1.25) {$\ell^{+}$};
\vertex [particle] (f2) at (2.5,0.75) {$\nu$};
\vertex [particle] (f3) at (2.5,0.25) {$b_{1}$};
\vertex [particle] (f4) at (2.5,-0.25) {$q_{1}$};
\vertex [particle] (f5) at (2.5,-0.75) {$\bar{q}_{2}$};
\vertex [particle] (f6) at (2.5,-1.25) {$\bar{b}_{2}$};

\vertex (v1) at (-1.5,0);
\vertex (v2) at (-0.5,0);
\vertex (v3) at (0.5,0.5);
\vertex (v4) at (1.5,1);
\vertex (v5) at (1.5, -0.5);
\vertex (v6) at (0.5,-0.5);

\propag [fer] (i1) to (v1);
\propag [antfer] (i2) to (v1);
\propag [boson] (v1) to [edge label=$Z^{\prime}$] (v2);
\propag [fer] (v2) to [edge label=$t$] (v3);
\propag [boson] (v3) to [edge label=$W^{+}$] (v4);
\propag [fer] (v3) to (f3);
\propag [antfer] (v4) to (f1);
\propag [fer] (v4) to (f2);
\propag [antfer] (v2) to [edge label=$\bar{t}$] (v6);
\propag [boson] (v6) to [edge label=$W^{-}$] (v5);
\propag [antfer] (v6) to (f6);
\propag [fer] (v5) to (f4);
\propag [antfer] (v5) to (f5);

\node at (-0.2,-1.3) [font=\fontsize{0.9em}{0.9em}\selectfont] {(b) $Z^{\prime}$ signal process};

\end{feynhand}
\end{tikzpicture}

\end{minipage}
\begin{minipage}{0.33\hsize}
\centering
\begin{tikzpicture}[scale=0.85]
\begin{feynhand}

\vertex [particle] (i1) at (-2.5,1.25) {};
\vertex [particle] (i2) at (-2.5,-1.25) {};

\vertex [particle] (f1) at (2.5,1.25) {$\ell^{+}$};
\vertex [particle] (f2) at (2.5,0.75) {$\nu$};
\vertex [particle] (f3) at (2.5,0.25) {$b_{1}$};
\vertex [particle] (f4) at (2.5,-0.25) {$q_{1}$};
\vertex [particle] (f5) at (2.5,-0.75) {$\bar{q}_{2}$};
\vertex [particle] (f6) at (2.5,-1.25) {$\bar{b}_{2}$};

\vertex (v1) at (-1.5,0);
\vertex (v2) at (-0.5,0);
\vertex (v3) at (0.5,0.5);
\vertex (v4) at (1.5,1);
\vertex (v5) at (1.5, -0.5);
\vertex (v6) at (0.5,-0.5);

\propag [gluon] (i1) to (v1);
\propag [gluon] (i2) to (v1);
\propag [gluon] (v1) to (v2);
\propag [fer] (v2) to [edge label=$t$] (v3);
\propag [boson] (v3) to [edge label=$W^{+}$] (v4);
\propag [fer] (v3) to (f3);
\propag [antfer] (v4) to (f1);
\propag [fer] (v4) to (f2);
\propag [antfer] (v2) to [edge label=$\bar{t}$] (v6);
\propag [boson] (v6) to [edge label=$W^{-}$] (v5);
\propag [antfer] (v6) to (f6);
\propag [fer] (v5) to (f4);
\propag [antfer] (v5) to (f5);

\node at (-0.2,-1.3) [font=\fontsize{0.9em}{0.9em}\selectfont] {(c) $t\bar{t}$ background process};
\end{feynhand}
\end{tikzpicture}
\end{minipage}
\caption{\label{fig:feyn} Examples of Feynman diagram for the signal and background processes.}
\end{figure}
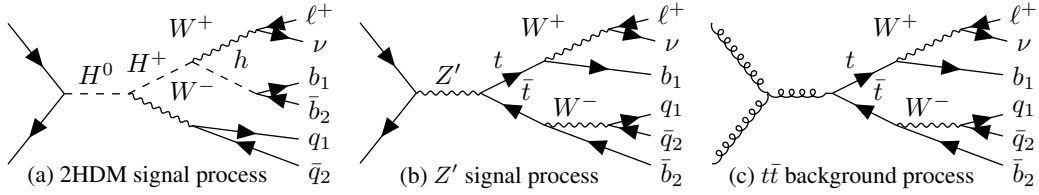

The training data were produced using particle physics simulations: proton--proton collision events were generated by MadGraph5\_aMC@NLO~\cite{Alwall:2014hca} at a center of mass energy of 13~TeV, with showering and hadronization performed by Pythia8~\cite{Sjostrand:2014zea} and detector response simulated by Delphes~\cite{Selvaggi_2014}. In this study, two types of datasets are defined as follows:

\begin{itemize}
\setlength{\leftskip}{-0.5cm}
\item 2HDM dataset: Two-Higgs-doublet model (2HDM)~\cite{2HDM, 2HDM2}, which introduces additional Higgs bosons, $H^{0}$, $A$ and $H^{\pm}$, is used as the signal event. Top pair production ($t\bar{t}$) of the Standard Model is used as the background event. Figure~\ref{fig:feyn}~(a) and (c) show Feynman diagrams for the signal and background processes. The final state particles are one lepton~($\ell$, electron or muon), one neutrino~($\nu$), two $b$-quarks~($b$), and two light-quarks~($q$). 
\item $Z^{\prime}$ dataset: A heavy neutral particle ($Z^{\prime}$) decaying into top quark pair~\cite{Altarelli:1989ff, Fuks2017ACF} is used as the signal event. The background events are the $t\bar{t}$ process. The decay chains of the top pairs are the same between signal and background events, as shown in Figure~\ref{fig:feyn}~(b) and (c).
\end{itemize}

Neutrinos were reconstructed as a missing transverse momentum object. $b$-quark and light-quark were reconstructed as $b$-jet and light-jet objects, respectively. In this study, the object-type is indicated by an integer, and the four-momenta of each object ($p_{\rm T}$, $\eta$, $\phi$, mass) and object-type were employed as input variables. The log transformation was applied to $p_{\rm T}$ and mass to fit the values within a reasonable range. 5~$\times$~10$^{5}$, 5~$\times$~10$^{4}$, and 5~$\times$~10$^{4}$ events were independently generated for training, validation, and test phases for each signal and background process. 

\subsection{Definition of auxiliary task labels}
\begin{figure}[htbp]
  \begin{minipage}[b]{0.33\linewidth}
    \centering
    \includegraphics[clip, width=0.9\columnwidth]{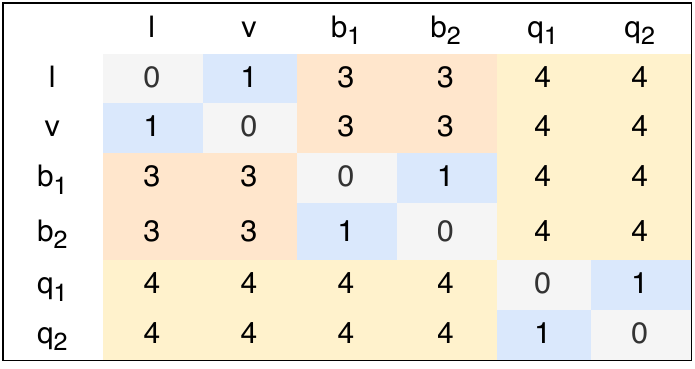}
    \subcaption{2HDM signal process}
  \end{minipage}
  \begin{minipage}[b]{0.33\linewidth}
    \centering
    \includegraphics[clip, width=0.9\columnwidth]{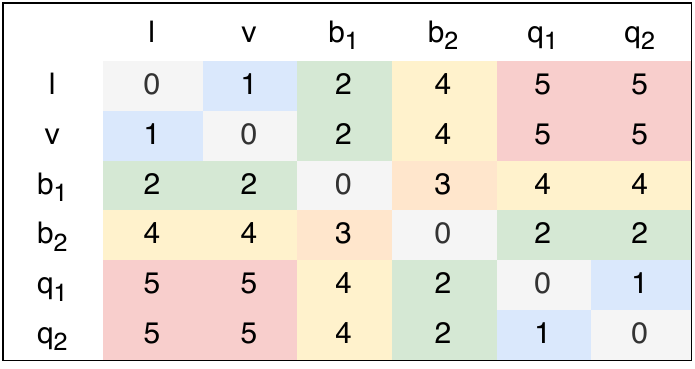}
    \subcaption{$Z^{\prime}$ signal process}
  \end{minipage}
    \begin{minipage}[b]{0.33\linewidth}
    \centering
    \includegraphics[clip, width=0.9\columnwidth]{zprime_label.pdf}
    \subcaption{$t\bar{t}$ background process}
  \end{minipage}
  \caption{\label{fig:label} Examples of the label matrices for each process corresponding to Figure~\ref{fig:feyn}. There are s- and t-channels in the $t\bar{t}$ process; however, the t-channel uses the same label as the s-channel to fit the same class range with the other processes. The max class label is 5 in this study.}
\end{figure}

The label of the auxiliary task is defined by counting the particle decay points, including gluon splitting points, between observed objects. Figure~\ref{fig:label} presents examples of the label matrices corresponding to Figure~\ref{fig:feyn}. For instance, the label between $\ell$ and $\nu$ in the $t\bar{t}$ process is 1 because they originated from the same $W$ decay; moreover, the label between $\ell$ and $b_{1}$ is 2 because there are two decay points: $t$ and $W$ decays. The information of truth particles in Delphes outputs is analyzed event by event to associate the observed objects with the truth particles to determine these labels.
Although many diagrams, including the next-to-leading order diagram, can be considered for a given physics process, we specifically assume the tree-level diagrams in Figure~\ref{fig:feyn} to prepare the label matrices in this study. The label matrices are represented by an edge-level class in the GNN. 

\section{Proposed decay-aware neural network}
\begin{figure}[htbp]
 \includegraphics[width=1.0\textwidth]{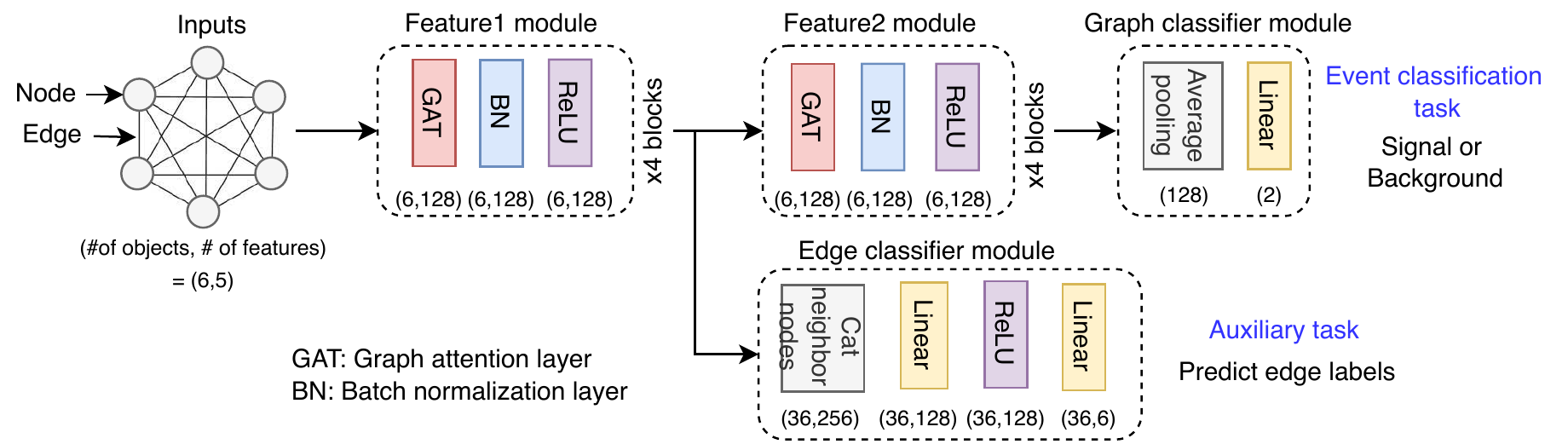}
 \caption{\label{fig:model} Overview of the proposed DL model. The numbers in the bracket indicate output shapes.}
\end{figure}

Figure~\ref{fig:model} shows an overview of the proposed decay-aware neural network. A graph structure composed of nodes and edges was built for the input data. In this study, each node in the graph corresponds to an object. Thus, object features: four-momentum and object-type are assigned to the node attribute. Edges are prepared as a fully-connected bidirectional graph, including a self-loop.

The model consisted of four parts: feature1, feature2, edge classifier, and graph classifier modules. Each feature module consists of a stack of blocks, where each block consists of the graph attention network (GAT) layer~\cite{brody2022how, NIPS2017_3f5ee243}, batch normalization (BN) layer~\cite{10.5555/3045118.3045167}, and ReLU activation function. The feature1 module is a shared representation of the concept of multi-task learning. The graph classifier module consists of an average pooling, which averages node attributes, and a fully connected layer (Linear) to predict the graph-level class, that is, a signal or background event. The edge classifier module is aimed at predicting the edge-level class, which was described in Section~3.1. Edge features are obtained by concatenating the source and target node attributes in the edge classifier module; then, they are processed by Linear, ReLU, and Linear layers. 

\section{Experiments}
Our proposed model for this experiment was implemented using PyTorch~\cite{NEURIPS2019_9015} and DGL~\cite{wang2019dgl} and is available at~\cite{hepdecay}. All executions used a local cluster of NVIDIA Tesla A100 graphics cards.

A loss function is defined as $L = L^{\rm graph} + \alpha \cdot L^{\rm edge}$, where $L^{\rm graph}$ and $L^{\rm edge}$ are loss functions for the event classification task and auxiliary task, respectively. \mbox{CrossEntropyLoss} and \mbox{BCEWithLogitsLoss} functions in PyTorch were used.  A parameter of $\alpha$ is introduced to scale the loss value of the auxiliary task. Therefore, a normal event classification without the auxiliary task is performed if $\alpha = 0$. The best epoch for the validation data was used as the final weight parameters after the training was performed for up to 100 epochs to minimize $L$. The SGD~\cite{https://doi.org/10.48550/arxiv.1609.04747} algorithm was used as an optimizer, and the learning rate was decreased from 0.01 to 0.0001 by the cosine annealing algorithm~\cite{https://doi.org/10.48550/arxiv.1608.03983}. The batch size was fixed at 2,048. Other hyperparameters, such as the number of nodes in the GAT layer, were optimized by a grid search with the fixed parameter of $\alpha = 0$.

\begin{figure}[htbp]
  \begin{minipage}[b]{0.5\linewidth}
    \centering
    \includegraphics[clip, width=0.98\columnwidth]{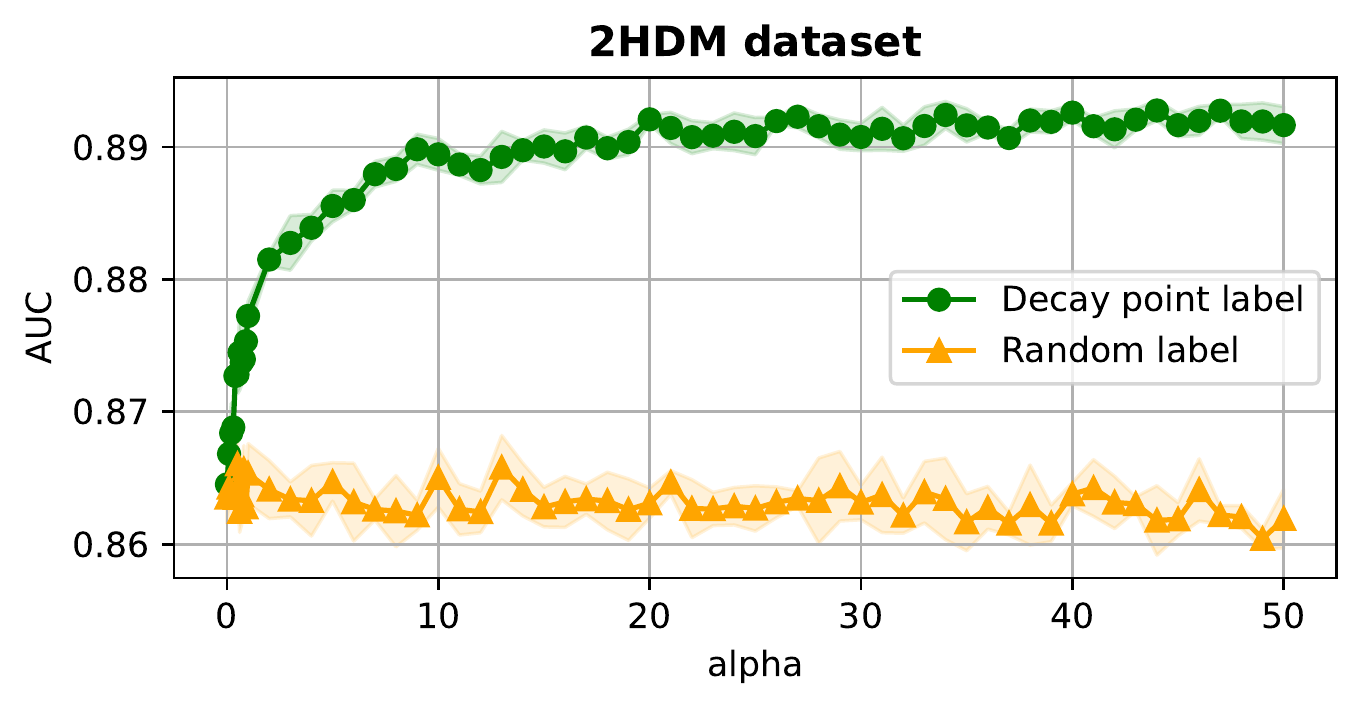}
  \end{minipage}
  \begin{minipage}[b]{0.5\linewidth}
    \centering
    \includegraphics[clip, width=1\columnwidth]{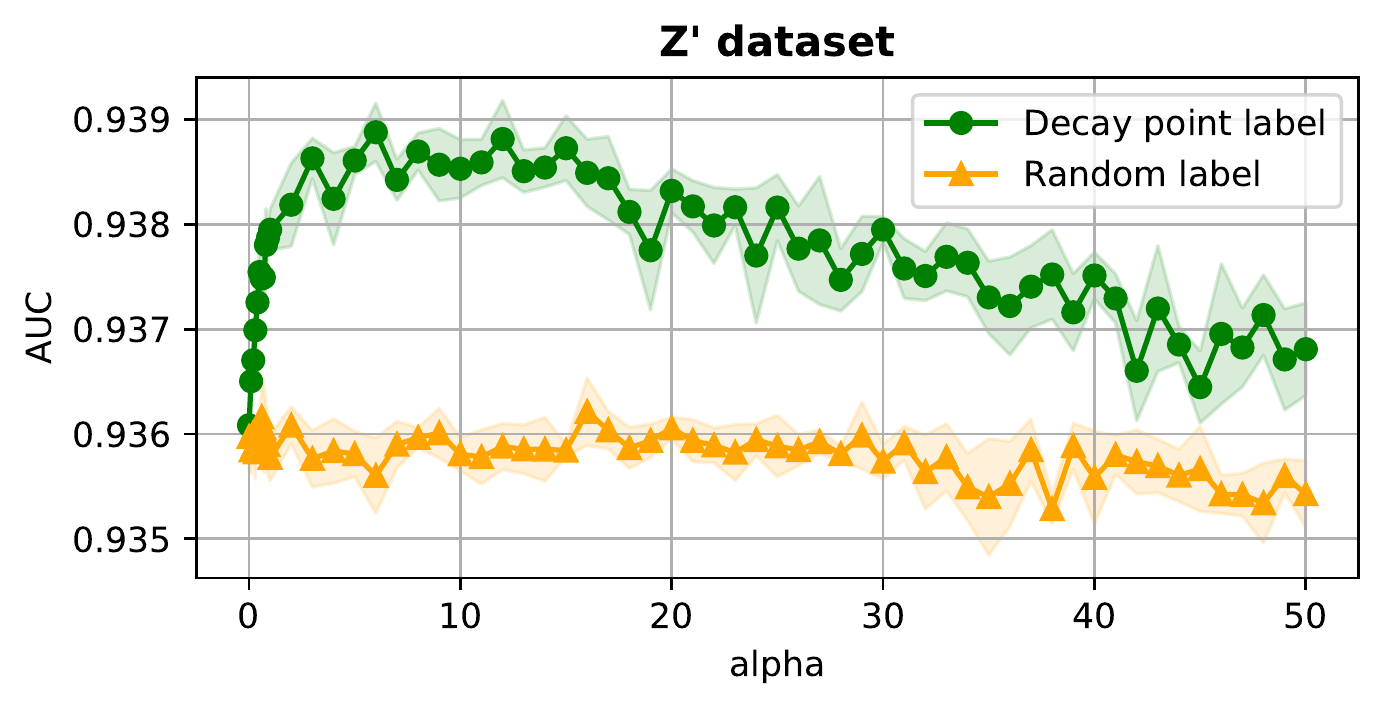}
  \end{minipage}
    \caption{\label{fig:auc} AUC of the event classification task for each dataset. The AUC values are an average of 5 runs with different random seeds for the test data. The error bands show one standard deviation.}
\end{figure}

Figure~\ref{fig:auc} presents obtained AUC values of the event classification task in terms of $\alpha$. The green points show the values using the auxiliary task labels described in Section~3.1. The figures confirm that the AUCs are improved by adding the auxiliary task, that is, $\alpha$ values are greater than~0. The best $\alpha$ values were 44 (2HDM dataset) and 6 ($Z^{\prime}$ dataset). The orange points show the values using random labels in the auxiliary task. No improvements were observed by the random labels as expected. The AUC values of the auxiliary task with the best $\alpha$ values were 0.968 (2HDM dataset) and 0.995 ($Z^{\prime}$ dataset), which were obtained by averaging each class with weights.

BDT and simple MLP models were also executed to compare the performance. Table~\ref{table:auc} summarizes observed AUCs for each experimental condition. The ``low-level" in the BDT and MLP means that the same input variables with our model, that is, the four-momenta and object-type for each object, were used with a one-dimensional shape. The ``w/ high-level" means that the invariant masses of $m_{bb}$, $m_{\ell\nu bb}$, $m_{\ell\nu qq}$, and $m_{bbWW}$ were calculated from the observed objects, and were added to the input variables to utilize the high-level features. The BDT was executed using XGBoost~\cite{Chen:2016:XST:2939672.2939785}. The MLP model consisted of a stack of Linear, BN layers, and ReLU activation functions. The max depth in BDT and the number of layers and nodes in MLP were optimized by a grid search. The table confirms that our model with the best $\alpha$ value shows better performances for both 2HDM and $Z^{\prime}$ datasets. In comparison to the BDT and simple MLP models, our proposed model can extract high-level features effectively based on particle decay information. 
\begin{table}[h]
 \caption{AUC values of the event classification task for each experimental condition. The AUC values are an average of 5 runs with different random seeds for the test data. The errors show one standard deviation. The max depth in BDT is 5 (2HDM) and 4 ($Z'$). The number of Linear layers in MLP is 5 (2HDM) and 4 ($Z'$). The number of nodes in each Linear layer is 64 for both datasets.}
 \label{table:auc}
 \centering
  \begin{tabular}{ccccccc}\hline\hline
 \multirow{2}{*}{2HDM}  &\multicolumn{2}{c}{Our model}      &\multicolumn{2}{c}{BDT}& \multicolumn{2}{c}{MLP} \\
  \cmidrule(lr){2-3} \cmidrule(lr){4-5} \cmidrule(lr){6-7}
  &$\alpha=0$&best $\alpha$ value&low-level & w/ high-level& low-level& w/ high-level \\\hline\hline
\multirow{2}{*}{2HDM} & 0.865             &{\bf 0.893}                  & 0.794         & 0.838               & 0.862 & 0.865 \\
                                    & $\pm$ 0.002  &  {\bf $\pm$ 0.001} &  $\pm$ 0.001 &  $\pm$ <0.001 &  $\pm$ 0.001 &  $\pm$ 0.001 \\\hline
\multirow{2}{*}{$Z^{\prime}$} & 0.936            &{\bf 0.939}              & 0.915             & 0.922            & 0.932             & 0.933 \\
                                              & $\pm$ <0.001  &  {\bf $\pm$ <0.001} &  $\pm$ <0.001 &  $\pm$ <0.001 &  $\pm$ <0.001 &  $\pm$ <0.001 \\\hline\hline
    \end{tabular}
\end{table}

Our experiments have the following limitation. The training of our model is much slower than the MLP model, which was approximately 36 batch/s~(our model) and 300 batch/s~(MLP model). Due to this computational constraint, we were unable to increase the signal types and the number of training events as we had intended. Additionally, we were interested in conducting a more thorough analysis of the edge-level outputs, such as the attention weights in the GAT layer, to increase explainability. Therefore, improving the training speed and explainability is a future subject.

\section{Conclusion}
In this paper, the decay-aware neural network based on the multi-task learning technique was discussed. The label of auxiliary task is defined based on the particle decay information to introduce an inductive bias with our domain knowledge. We successfully trained the event classification task and the auxiliary task simultaneously using the GNN architecture. Our experiments confirmed that the event classification performance improved significantly by adding the auxiliary task, and  better performances were achieved compared with BDT and MLP models due to the inductive bias. 

\section*{Broader Impact}
This work can potentially contribute to the explainability of machine learning, which would be interested in a wide range of scientific communities. In the field of collider physics, researchers have a deep understanding of data based on particle physics theories and past experimental results. We can probe behaviors of the machine learning by introducing the domain knowledge as demonstrated in this paper. We believe that this approach is effective to understand what machine learning has learned and improve its explainability. This study is aimed at improving a problem in pure fundamental science, and we do not expect our study to result in a negative social impact.

\section*{Acknowledgment}
This work was supported by JSPS KAKENHI Grant Number JP22K14050 and Institute for AI and
Beyond of the University of Tokyo.

\bibliographystyle{junsrt}
\bibliography{refs}

\appendix



\end{document}